\documentclass[aps,prl,twocolumn,superscriptaddress,longbibliography]{revtex4-2}
\usepackage[T1]{fontenc}
\usepackage{lmodern}
\usepackage{amsmath,amssymb,bm,mathtools}
\usepackage{graphicx,microtype}

\newcommand{\dd}{\mathrm{d}}
\newcommand{\ii}{\mathrm{i}}
\newcommand{\Tr}{\operatorname{Tr}}
\newcommand{\avg}[1]{\left\langle #1\right\rangle}
\newcommand{\ket}[1]{\lvert #1\rangle}
\newcommand{\bra}[1]{\langle #1\rvert}
\newcommand{\bh}{{\bm h}}
\newcommand{\bM}{{\bm M}}

\begin{document}

\title{Dynamical Keldysh model for  colored noise and finite relaxation rate}

\author{Fatima Zohra Bendjeddou}
\affiliation{The Abdus Salam International Centre for Theoretical Physics, Strada Costiera 11, I-34151 Trieste, Italy}
\author{D. V. Efremov}
\affiliation{IFW Dresden, Helmholtzstr. 20, 01069 Dresden, Germany}
\author{M. N. Kiselev}
\affiliation{The Abdus Salam International Centre for Theoretical Physics, Strada Costiera 11, I-34151 Trieste, Italy}

\date{\today}

\begin{abstract}
The Keldysh model describes quantum dots subjected to Gaussian noise with infinite correlation time. For multi-component noise coupled through non-commuting matrices, the spectrum exhibits a pseudogap. We solve the model for coloured noise with any correlation time, $1/ \gamma$. Finite memory leads to the pseudogap being filled, its peaks merging and the line narrowing.
\end{abstract}

\maketitle

Electrons in quantum dots and other gated nanostructures are subject to
electric noise from the gates and their environment, with correlation times
spanning many orders of magnitude~\cite{Paladino2014,Burkard2023}.
Two limiting regimes are typically considered: the slow-noise limit, in
which the noise correlation time greatly exceeds the timescale of the
electron dynamics, and the fast (Markovian) limit, in which it is much
shorter.
In the fast-noise limit the noise is effectively white on the timescale of
the electron dynamics, and the problem reduces to pure dephasing: each
level acquires a Lorentzian broadening.
In the opposite limit of infinite noise correlation time, the noise is
frozen on the timescale of the electron dynamics, and the ensemble average
is performed over static realisations of the random field.
In particular, for a Gaussian random field the single quantum dot problem maps
onto the Keldysh model,
which admits an exact diagrammatic
solution~\cite{Keldysh1965,Efros1971,SadovskiiDiagrammatics,Kikoin2012,Kiselev2009,Efremov2022,Kiselev2026b}.

\begin{figure}[t]
\centering
\includegraphics[width=\columnwidth]{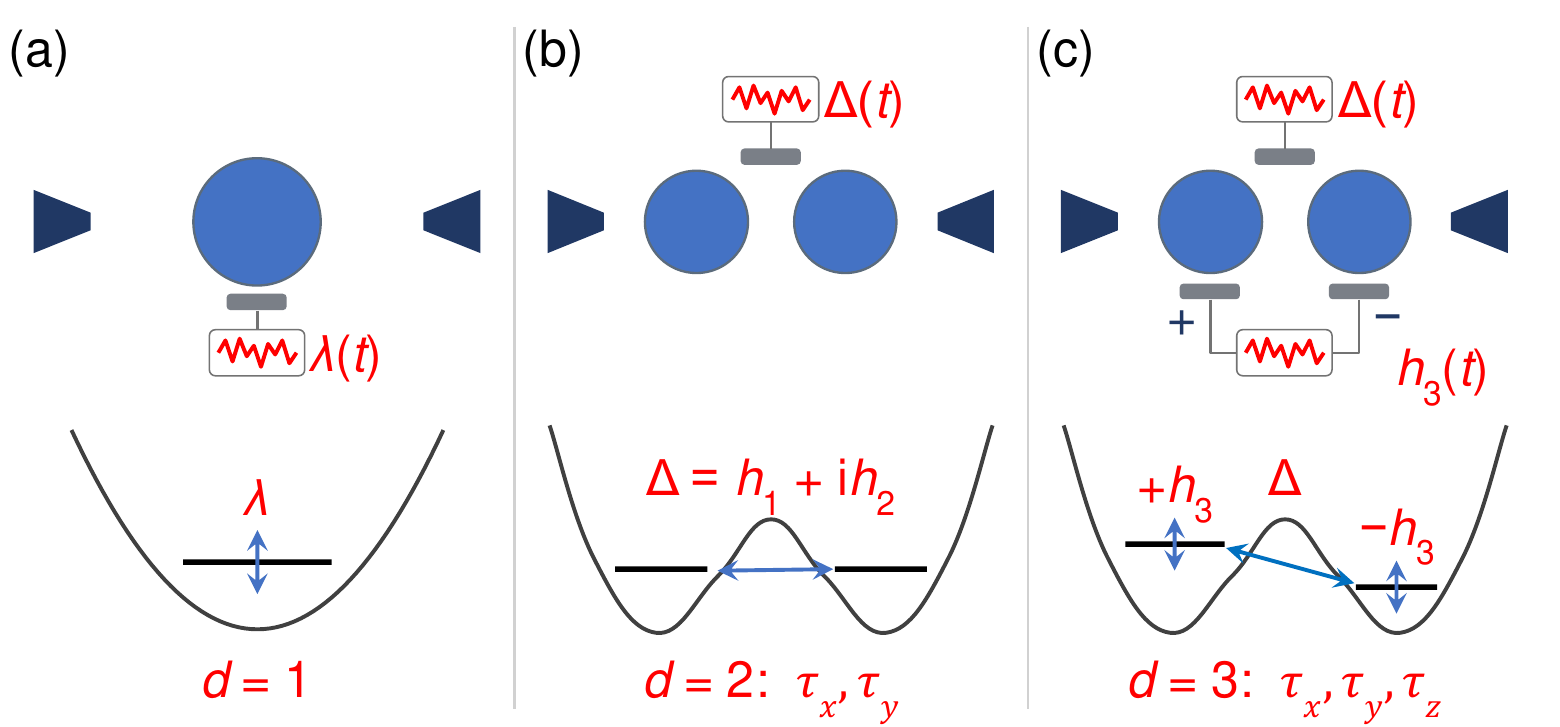}
\caption{Gate-defined quantum dots as realizations of the $d$-component
dynamical Keldysh model (top) and their confining potentials and levels
(bottom).
(a)~A noisy gate shifts the level of a single dot by $\lambda(t)=h_1(t)$, $d=1$.
(b)~Barrier noise in a double dot makes the tunneling amplitude
$\Delta=h_1+\ii h_2$ fluctuate, $d=2$, with vertices $\tau_{x,y}$.
(c)~Antisymmetric noise $\pm h_3(t)$ on the plunger gates adds the vertex
$\tau_z$, $d=3$.}
\label{fig:dots}
\end{figure}

Coupled dots make the noise multicomponent~\cite{Kiselev2009,Efremov2022} as it is demonstrated in
Fig.~\ref{fig:dots}.
A noisy gate modulates the level of a single dot ($d=1$).
In a double dot, barrier noise renders the tunnel amplitude random ($d=2$),
and antisymmetric noise on the switching gates adds a third component
($d=3$).
For $d=2,3$ the noise components couple through anticommuting Pauli
matrices; a frozen field $\bh$ then splits the level into
$\varepsilon_0\pm|\bh|$, and the density of states (DOS) develops a
pseudogap $\propto|\varepsilon-\varepsilon_0|^{d-1}$~\cite{Kiselev2009,Efremov2022}.
The same Pauli-matrix structure underlies Sadovskii's models of the
pseudogap state, whose commensurate, incommensurate, and Heisenberg
versions correspond to $d=1,2,3$~\cite{Sadovskii1979,Schmalian1999,KuchinskiiSadovskii1999}.

However, real noise has a finite memory time. 
For $d=1$ the problem is Abelian and admits an exact solution in terms of
the time-evolution operator; it was first solved by Kubo and
Anderson~\cite{Kubo1954,Anderson1954,Kubo1962}.
For $d\geq2$ the noise vertices at different times do not commute, and the
time ordering cannot be rearranged as in the Abelian case.
Continued fractions with damping factors $\ii k\gamma$, constructed from
the diagram rules of Elyutin and Sadovskii, are natural candidates for
the solution~\cite{Sadovskii1979}.

In this article we present an exact solution of the $d$-component dynamical
Keldysh model for Gaussian noise with a finite correlation time.
The averaged evolution operator is generated by $d$ oscillator modes whose
quanta decay at rate $\gamma$.
The anticommutation of the noise vertices yields an exact continued fraction.
Its coefficients reproduce the commensurate, incommensurate, and Heisenberg
combinatorial factors, proving that Sadovskii-type fractions are exact in
the time domain.
Finite memory closes the pseudogap for any $\gamma>0$; the two side peaks
merge at $\gamma_c \sim W$.
In the slow-noise regime the pseudogap is smeared not on the scale
$\gamma$ but on the much larger scale $(\gamma W^2)^{1/3}$, and the DOS at
the band center grows as $(\gamma/W)^{(d-1)/3}$.

\begin{figure*}[t]
\centering
\includegraphics[width=2.0\columnwidth]{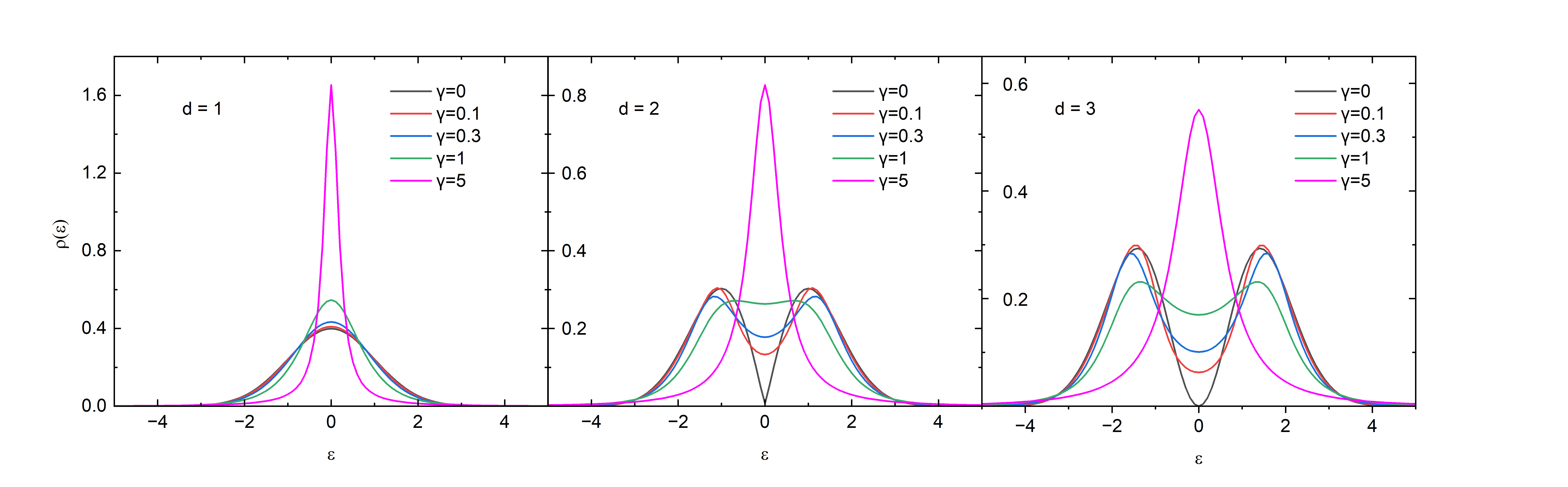}
\caption{Evolution of the DOS with the dimensionless correlation rate
$\gamma$ for $d=1,2,3$.
The curves show the continued-fraction solution at
$\alpha=0,\,0.3,\,1,\,5$.
For $d=2,3$ the pseudogap fills as the field becomes dynamical;
fast fluctuations ($\gamma\gg W$) produce a central peak. Parameters: $W =1$.}
\label{fig:merge}
\end{figure*}

\paragraph{Model.}
The electron occupies the levels $\mu$ of a dot cluster or a pseudospin and
couples to a $d$-component classical field $\bh(t)$,
\begin{equation}
 \hat H(t)=\sum_{\mu\nu}\big[\varepsilon_0\delta_{\mu\nu}+V_{\mu\nu}(t)\big]
 \hat c^\dagger_\mu\hat c^{}_\nu,\quad V(t)=\sum_{j=1}^{d}h_j(t)M_j.
 \label{eq:H}
\end{equation}
The Hermitian noise vertices anticommute, $\{M_j,M_k\}=2\delta_{jk}$, so that
$(\bh\cdot\bM)^2=|\bh|^2$; $M_1=1$ for $d=1$, and the $M_j=\tau_j$ are Pauli
matrices for $d=2,3$. In this article we consider the components of $\bh$ are independent stationary
Gaussian processes with zero mean and
\begin{equation}
 \avg{h_j(t)h_k(t')}=\delta_{jk}D(t-t'),\qquad D(t)=W^2e^{-\gamma|t|}.
 \label{eq:noise}
\end{equation}
This is the Ornstein--Uhlenbeck process, which is Markovian for any
$\gamma$ but non-Markovian dynamics for multi-component noise~\cite{Doob1942}: $W$ is the standard deviation of each component and
$1/\gamma$ its correlation time: 
\begin{equation}
    d \bh = - \gamma \bh dt + \sqrt{2\gamma} W d \mathbf{B}, \qquad \langle dB_i dB_j \rangle = \delta_{ij} dt.
\end{equation}
The Keldysh limit $\gamma\to0$ is a frozen
field, the stationary density is $P_d(\bh) = (2\pi W)^{-d/2} \exp[-|\bh|^2/(2W^2)]$. Our finite-time reduction assumes equal $W$ and equal $\gamma$ for all active components, zero mean-field and no deterministic level splitting.  

We study the retarded Green function per level,
\begin{equation}
 G_d(z)=-\frac{\ii}{n_d}\int_0^\infty\!\dd t\,e^{\ii zt}\,\Tr\avg{U(t)},
 \label{eq:G}
\end{equation}
where $U(t)=\mathcal T\exp[-\ii\int_0^t V(s)\,\dd s]$ is the time-ordered
evolution operator, $z=\varepsilon+\ii0^+$, and $n_d$ is the
dimension of the internal space ($n_1=1$, $n_2=n_3=2$). The DOS per level is
$\nu_d(\varepsilon)=-\operatorname{Im}G_d/\pi$. Below we use
$x=\varepsilon/W$, $\alpha=\gamma/W$, and
$\rho_d(x,\alpha)=W\nu_d$.

\paragraph{Scalar noise.}
For $d=1$ the noise vertices at different times commute, so the exponent
$X=-\ii\int_{0}^{t}h(s)\,\dd s$ is a Gaussian variable with zero mean.
Its average is then exact to the second cumulant,
$\avg{e^{X}}=\exp[\avg{X^{2}}/2]$~\cite{Kubo1962}, which gives
    \begin{align*}
 \avg{U(t)}&=\exp\Big[-\frac12\int_0^t\!\!\int_0^t\dd s\,\dd s'\,D(s-s')\Big]
 \\&=\exp\Big[-\frac{W^2}{\gamma^2}\big(\gamma t-1+e^{-\gamma t}\big)\Big].
\end{align*}
It gives Kubo's line
shape~\cite{Kubo1954,Anderson1954},
\begin{align}
 \Psi_1(\zeta,\alpha)&=-\ii\int_0^\infty\dd s\,
 \exp\Big[\ii\zeta s-\frac{\alpha s-1+e^{-\alpha s}}{\alpha^2}\Big]
 \label{eq:kubo}\\
 &=-\ii\,e^{1/\alpha^2}\alpha^{2p-1}\big[\Gamma(p)-\Gamma(p,\alpha^{-2})\big],
 \label{eq:kubo-gamma}
\end{align}
where $\Gamma(p,y)$ is the upper incomplete gamma function and $ p=\frac{1-\ii\alpha\zeta}{\alpha^2}$.  For $\alpha\to0$ the
exponent tends to $\ii\zeta s-s^2/2$; for
$\alpha\to\infty$ at fixed $\zeta$, $\Psi_1\to(\zeta+\ii/\alpha)^{-1}$, the
motionally narrowed line.

\paragraph{Axillary boson representation.}
For $d>1$ expanding $U(t)$ before averaging gives a time-ordered series whose $m$th term
contains the product $M_{j_1}\cdots M_{j_m}$ in the order of the times,
multiplied by the moment $\avg{h_{j_1}(t_1)\cdots h_{j_m}(t_m)}$. For Gaussian
noise each moment is a sum over all pairings~\cite{Isserlis1918}, nested and
crossed alike. For $d=1$ the series exponentiates into Kubo's line
shape~\cite{Kubo1962}; for $d\ge2$ the order of the matrices must be kept. We
reproduce the pairings with $d$ bosonic modes $a_j$,
$[a_j,a_k^\dagger]=\delta_{jk}$, and their number operator
$N=\sum_ja_j^\dagger a_j$. 

Since the matrices and the oscillators act in different spaces, the series
then sums to
\begin{align}
 \avg{U(t)}&=\bra0\mathcal T e^{-\ii W\int_0^t\dd s\sum_jM_jX_j(s)}\ket0
 \notag\\
 &=\bra0e^{-\ii\mathcal Ht}\ket0,\qquad
 \mathcal H=W(B+B^\dagger)-\ii\gamma N,
 \label{eq:Heff}
\end{align}
with $B=\sum_jM_ja_j$. The first line is the interaction-picture expansion of
$e^{-\ii\mathcal Ht}$ with respect to $-\ii\gamma N$. Consequently,
\begin{equation}
 G_d(z)=\frac{1}{n_d}\Tr\bra0(z-\mathcal H)^{-1}\ket0 .
 \label{eq:resolvent}
\end{equation}
The auxiliary modes keep the full time ordering, and the vacuum encodes the stationary distribution of the field at $t=0$. The same result follows from the stochastic Liouville equation of the Markovian field~\cite{Kubo1963}, in which $a_j$ and $a_j^\dagger$ are the
ladder operators of the Hermite eigenfunctions of the Ornstein--Uhlenbeck generator $-\gamma N$. Equation~\eqref{eq:resolvent} holds for any
Hermitian vertices.

\paragraph{Single chain.}
The oscillator commutators give \cite{SM}
\begin{equation}
 \{B,B^\dagger\}=2N+d,\qquad[N,B^\dagger]=B^\dagger,\qquad B\ket0=0.
 \label{eq:algebra}
\end{equation}
Let $\chi$ be an arbitrary internal spinor. The states
$\ket{f_k}=(B^\dagger)^k\ket{0,\chi}$ have $N=k$, and Eq.~\eqref{eq:algebra}
gives $B\ket{f_k}=v_d(k)\ket{f_{k-1}}$ with
$v_d(k)=2k+d-2-v_d(k-1)$ and $v_d(0)=0$, i.e.,
\begin{equation}
 v_d(k)=k+d-1\ \ (k\ \text{odd}),\qquad v_d(k)=k\ \ (k\ \text{even}).
 \label{eq:v}
\end{equation}
In the normalized states $\ket k$, $B^\dagger\ket{k-1}=\sqrt{v_d(k)}\,\ket k$
and $B\ket{k}=\sqrt{v_d(k)}\,\ket{k-1}$. The chain $\{\ket k\}$ is thus
invariant under $\mathcal H$, which is tridiagonal in it with diagonal
elements $-\ii k\gamma$ and off-diagonal elements $W\sqrt{v_d(k)}$. No internal
matrix survives, and Eq.~\eqref{eq:resolvent} becomes \cite{Schmalian1999}
\begin{equation}
 G_d(z)=\cfrac{1}{z-\cfrac{dW^2}{z+\ii\gamma-\cfrac{2W^2}{z+2\ii\gamma
 -\cfrac{(d+2)W^2}{z+3\ii\gamma-\cdots}}}}\,.
 \label{eq:cf}
\end{equation}
Equations~\eqref{eq:algebra}--\eqref{eq:cf} hold for any number $d$ of
anticommuting vertices. For $d=1$, Eq.~\eqref{eq:cf} is the continued fraction
of Kubo's line shape~\cite{Kuchinskii2024}. 
 
\begin{figure}[t]
\centering
\includegraphics[width=0.8\columnwidth]{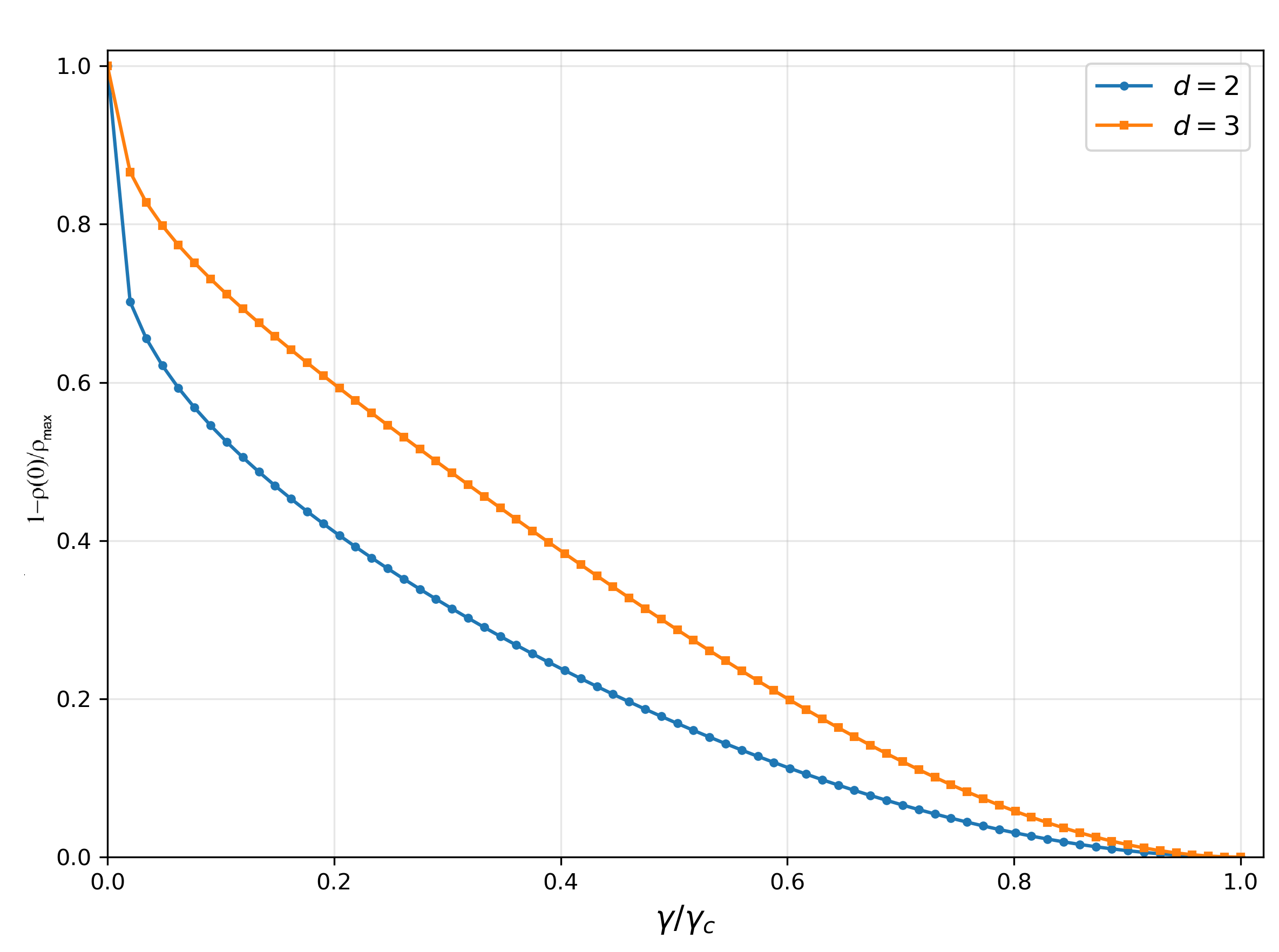}
\caption{Relative pseudogap depth for the $d=2$ and $d=3$ as a
    function of the rescaled correlation rate $\gamma/\gamma_c$.}
\label{fig:scaling1}
\end{figure}

\paragraph{Density of states.}
For $\gamma=0$ the DOS is the radial distribution of
$|\bh|$~\cite{Kiselev2009,Efremov2022},
\begin{equation}
 \rho_d(x,0)=\frac{|x|^{d-1}e^{-x^2/2}}{2^{d/2}\Gamma(d/2)},
 \label{eq:static}
\end{equation}
a Gaussian for $d=1$ and a linear ($d=2$) or quadratic ($d=3$) pseudogap with
maxima at $x=\pm\sqrt{d-1}$.

\paragraph{Filling and merging at finite $\gamma$.}
Any $\gamma>0$ fills the pseudogap. At $\alpha=0.1$ the central DOS is already
$\rho_2(0)=0.133$ and $\rho_3(0)=0.063$, compared with peak heights of about
$0.3$. As $\alpha$ grows, the side peaks first move outward and then merge,
Fig.~\ref{fig:merge}. The central curvature changes sign at
\begin{equation}
 \alpha_c^{(2)}=1.2485,\qquad\alpha_c^{(3)}=1.9010.
 \label{eq:alphac}
\end{equation}
Near $\alpha_c$ the
peak position and the relative depth vanish as
$x_{\max}\sim  (\alpha_c-\alpha)^{1/2}$ and
$1-\rho(0)/\rho_{max}\sim (\alpha_c-\alpha)^2$ as one can see in Fig. \ref{fig:scaling1}. This smooth change of the line shape,
not a phase transition, is the matrix analog of the exchange narrowing of two
magnetic-resonance lines~\cite{Anderson1954,Kubo1954}: when $\gamma$ becomes
comparable with the typical splitting $2W\sqrt{d-1}$, the electron no longer
resolves the levels $\varepsilon_0\pm|\bh|$.

\paragraph{Fast noise.}
For $\alpha\gg1$ the first level of the chain dominates, and in the central
region $|x|\ll\alpha$
\begin{equation}
 \rho_d(x,\alpha)\simeq\frac{1}{\pi}\,\frac{d/\alpha}{x^2+(d/\alpha)^2},
 \label{eq:lorentz}
\end{equation}
a Lorentzian of half-width $dW^2/\gamma$: faster noise narrows the line, as for
$d=1$~\cite{Kubo1954,Anderson1954}. The far tails are not Lorentzian. For any
finite $\alpha$ and $|x|\to\infty$ the fraction gives
\begin{equation}
 \rho_d(x,\alpha)\simeq\frac{d\alpha}{\pi x^4},\qquad
 \nu_d(\varepsilon)\simeq\frac{d\,D(\varepsilon-\varepsilon_0)}
 {2\pi(\varepsilon-\varepsilon_0)^2},
 \label{eq:tail}
\end{equation}
with the noise spectrum $D(\omega)=2\gamma W^2/(\omega^2+\gamma^2)$: the tail is
the golden-rule response of the level to the noise at frequency
$\varepsilon-\varepsilon_0$~\cite{SM}. These tails make all moments beyond the
second diverge, so frozen-field sum rules do not apply at finite $\gamma$.
Holding the white-noise strength $D_0=W^2/\gamma$ fixed as $\gamma\to\infty$
gives $G_d=(z+\ii dD_0)^{-1}$.

\begin{figure}[t]
\centering
\includegraphics[width=\columnwidth]{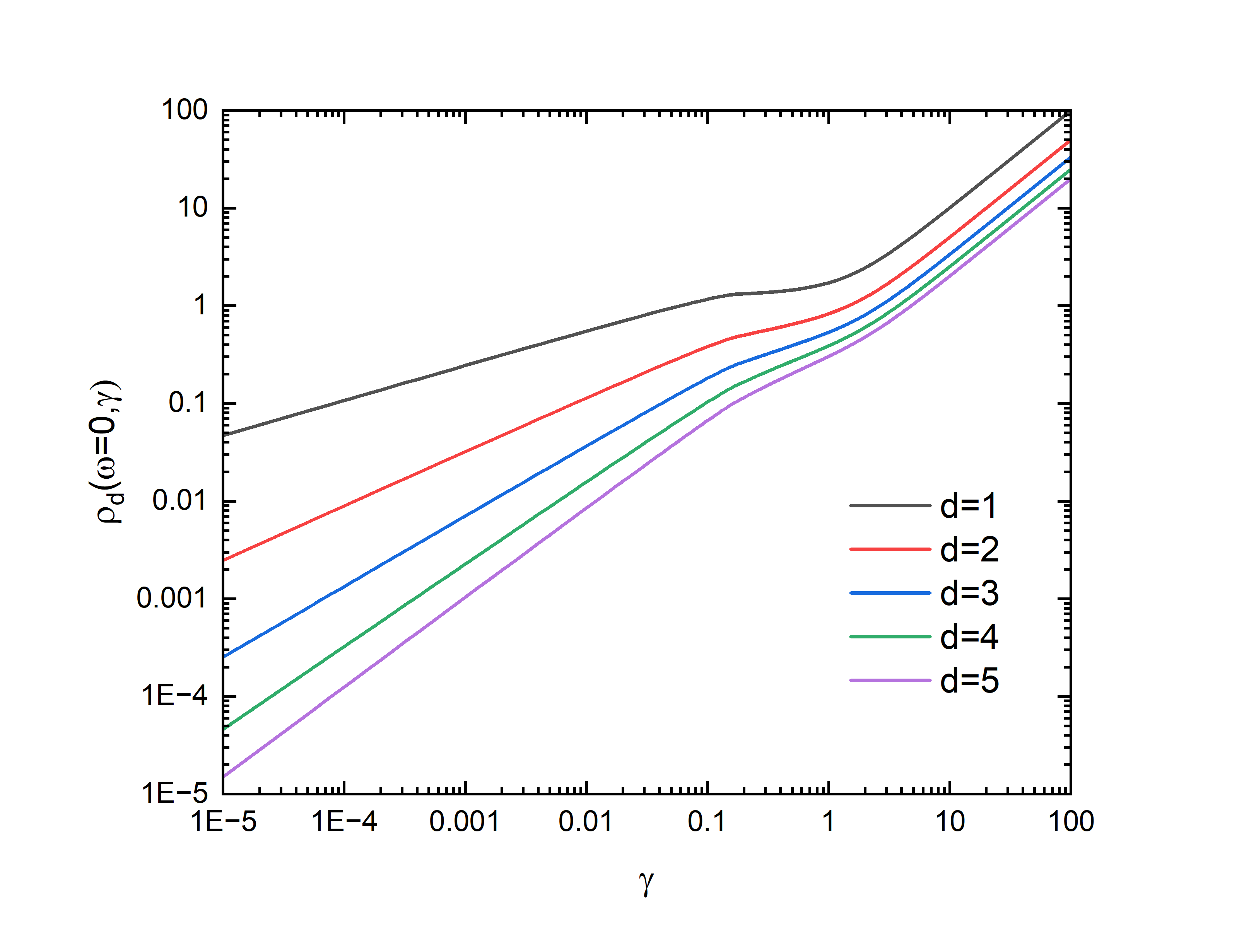}
\caption{Filling of the pseudogap by slow noise. Central DOS
$\rho_d(0,\alpha)$ from the continued fraction. For
$d=2,3$ the pseudogap fills with the fractional powers $\alpha^{1/3}$ and
$\alpha^{2/3}$; for $\alpha\gg1$ all curves approach the fast-noise value
$\alpha/(\pi d)$.}
\label{fig:scaling}
\end{figure}

\paragraph{Slow noise.}
For $\alpha\to0$ the spectrum approaches Eq.~\eqref{eq:static} away from the
center, but the filling of the pseudogap is singular (see Fig. \ref{fig:scaling}). The exact zero-energy
chain gives~\cite{SM}
\begin{equation}
	\begin{split}
		&\rho_d(0,\alpha)\simeq\mathcal C_d\,\alpha^{(d-1)/3},\\
		&\mathcal C_d=\frac{\Gamma\big(\frac d2\big)\,\Gamma\big(\frac{4-d}{6}\big)}
		{\sqrt2\,\pi\,\Gamma\big(\frac{d+2}{6}\big)}
		\Big(\frac{\sqrt2}{3}\Big)^{(d-1)/3},
	\end{split}
	\label{eq:slow}
\end{equation}
with $\mathcal C_2=0.3466$ and $\mathcal C_3=0.5958$; for $d=1$,
$\mathcal C_1=1/\sqrt{2\pi}$ is the static value. The fractional exponent
signals a new energy scale. During the time $1/E$ an electron near
$\varepsilon_0$ sees the field diffuse by $\delta h\sim W(\gamma/E)^{1/2}$,
which exceeds $E$ below $E_*=(\gamma W^2)^{1/3}$. The pseudogap is therefore
smeared on the scale $E_*\gg\gamma$, and the radial weight $|\bh|^{d-1}$ at
$|\bh|\sim E_*$ gives $\rho_d(0)\propto(E_*/W)^{d-1}=\alpha^{(d-1)/3}$. In the
chain, $E_*$ corresponds to the level $k_*\sim\alpha^{-2/3}$. Below it the even
and odd components of the zero-energy solution decouple, and they are matched
to the decaying solution through the Bessel functions $K_{(4-d)/6}$ and
$K_{(d+2)/6}$~\cite{SM}. The continued fraction approaches
Eq.~\eqref{eq:slow} with relative deviations of $3\times10^{-5}$ ($d=2$) and
$3\times10^{-3}$ ($d=3$) at $\alpha=10^{-8}$; see Fig.~\ref{fig:scaling}. Slow
noise thus fills the pseudogap efficiently: at $\gamma=10^{-3}W$ the central DOS
for $d=2$ already amounts to $11\%$ of the static peak height.

\paragraph{Conclusions.}
We have mapped the averaged Green function of a level in Gaussian Markovian
noise onto the vacuum resolvent of damped oscillators,
Eq.~\eqref{eq:resolvent}, and shown that anticommuting noise vertices reduce it
to a single exact continued fraction, Eq.~\eqref{eq:cf}. This settles the
status of Sadovskii-type fractions in the time domain and gives the complete
evolution of the pseudogap from the frozen to the white-noise limit.
Tunneling spectroscopy of weakly coupled dots (Fig.~\ref{fig:dots}) probes
$\nu_d(\varepsilon)$, so the line shape measures the noise correlation time:
two peaks near $\pm W\sqrt{d-1}$ for slow noise, a single central peak beyond
$\gamma_c$, and a motionally narrowed Lorentzian for $\gamma\gg W$. 
\paragraph{Note added.} After this work was completed, we became aware of Ref. \cite{Sadovskii2026b}, in which Sadovskii considered a closely related problem.
\bibliography{references}

\end{document}